\documentclass[fleqn,10pt,twocolumn]{wlscirep}
\usepackage[utf8]{inputenc}
\usepackage[T1]{fontenc}

\newcommand{\orcid}[1]{\href{https://orcid.org/#1}
{\includegraphics[width=10pt]{orcid.pdf}}}

\title{Evidence of Hadronic Emission from the brightest-of-all-time GRB 221009A}
\author[1]{Kai Wang}
\author[2]{Qing-Wen Tang}
\author[3,4]{Yan-Qiu Zhang}
\author[3,4]{Chao Zheng}
\author[3]{Shao-Lin Xiong}
\author[5]{Jia Ren}
\author[6]{Bing Zhang}
\affil[1]{Department of Astronomy, School of Physics, Huazhong University of Science and Technology, Wuhan 430074, China, Email: kaiwang@hust.edu.cn}
\affil[2]{Department of Physics, School of Physics and Materials Science, Nanchang University, Nanchang 330031, China, Email: qwtang@ncu.edu.cn}
\affil[3]{Key Laboratory of Particle Astrophysics, Institute of High Energy Physics, Chinese Academy of Sciences, Beijing 100049, China}
\affil[4]{University of Chinese Academy of Sciences, Chinese Academy of Sciences, Beijing 100049, China}
\affil[5]{School of Astronomy and Space Science, Nanjing University, Nanjing 210023, China}
\affil[6]{Nevada Center for Astrophysics and Department of Physics and Astronomy, University of Nevada, Las Vegas NV 89154, USA, Email: bing.zhang@unlv.edu}


\begin{abstract}
Acceleration of hadrons in relativistic shocks has been long expected and invoked to model GRB high-energy photon and neutrino emissions. However, so far there has been no direct observational evidence of hadronic emission from GRBs. The B.O.A.T. (``brightest of all time'') gamma-ray burst (GRB) 221009A had extreme energies (with an isotropic energy exceeding $10^{55}$ erg) and was detected in broad-band including the very-high-energy (VHE, $>100\,\rm GeV$) band up to $>10$ TeV. Here we perform a 
comprehensive spectral analysis of the GRB from keV to TeV energy range and perform detailed spectral and light curve modelings considering both the traditional synchrotron self-Compton process and the electromagnetic (EM) cascade process initiated by hadronic interactions by accelerated cosmic rays in the external shock. 
We find that the leptonic scenario alone is not adequate to account for the observations, whereas the proposed scenario with the combination of hadronic and leptonic components can well reproduce the multi-wavelength spectra and the light curve. This result reveals the existence of the accelerated hadronic component in the early afterglow of this extreme burst. According to this scenario, the observed TeV light curve should contain imprints of the prompt MeV emission.
\end{abstract}

\begin{document}

\flushbottom
\maketitle

%\newpage
%\thispagestyle{empty}

%\section{Introduction}\label{sec:introduction}

%Energetic Gamma Ray Experiment Telescope (EGRET)

% \textcolor{cyan}{\textbf{Introduction.}} 

Recently, very-high-energy (VHE, $>100\,\rm GeV$) gamma-rays have been detected during the Gamma-ray bursts (GRBs) afterglow decaying phase by imaging atmospheric Cherenkov telescopes (IACTs), e.g., GRB 190114C~\cite{2019Natur.575..455M,2019Natur.575..459M} and GRB 201216C~\cite{2022icrc.confE.788F} by The Major Atmospheric Gamma Imaging Cherenkov (MAGIC) observatory, GRB 180720B~\cite{2019Natur.575..464A} and GRB 190829A~\cite{2021Sci...372.1081H} by the High Energy Stereoscopic System (HESS) observatory. Very recently and unprecedentedly, a considerable amount of VHE photons with energy extending above $10\,\rm TeV$ were detected by the Large High Altitude Air Shower Observatory (LHAASO) from the B.O.A.T (“brightest of all time”) GRB 221009A during its very early stage when the GRB prompt emission is still lasting \cite{2022GCN.32677....1H,doi:10.1126/science.adg9328}. GRB 221009A provides a unique opportunity to study the origin of high-energy and VHE gamma-rays, revealing the physical properties of the GRB jet \cite{2023ApJ...947...53R,2023ApJ...947L..14Z,2023MNRAS.522L..56S,2023ApJ...942L..30S,2023ApJ...944L..34R,2023arXiv230211111W}.

GRBs are considered promising candidates to accelerate particles to ultrahigh-energy cosmic rays (UHECRs) by either internal shocks \cite{1995PhRvL..75..386W,WAXMAN200646} or external shocks \cite{1995ApJ...453..883V,Wick:2004vkr,Dermer_2006,Wang_2008,2008PhRvD..78b3005M}. High-energy ($>100\,\rm MeV$) and VHE gamma rays can be produced via hadronic processes, e.g. the photomeson production process ($p\gamma\to (p/n)\pi^0\pi^+\pi^-$) and the Bethe-Heitler process (BH, $p\gamma\to pe^+e^-$), between the accelerated cosmic rays and the prompt or afterglow electromagnetic (EM) emissions \cite{2009ApJ...705L.191A,2010ApJ...725L.121A,2012ApJ...746..164M,2012ApJ...757..115A,2018ApJ...857...24W,2021tang,2023ApJ...950...28R,2023arXiv230211111W}. A competitive model for the origins of high-energy and VHE gamma rays is the leptonic scenario, which is generally attributed to synchrotron radiation or the inverse Compton (IC) scattering of the low-energy photon field through energetic electrons accelerated by shocks ~\cite{2001ApJ...559..110Z,2009MNRAS.400L..75K,2007MNRAS.380...78G,2008MNRAS.385.1461Y,2010MNRAS.409..226K,2009A&A...498..677B,2011ApJ...739..103A,2011ApJ...729..114A,2012ApJ...757..115A,2013ApJ...773L..20L,2013ApJ...771L..33W,2014ApJ...788...36B,2017ApJ...844...92F,2019ApJ...884..117W}. Besides, the possible hybrid scenario including the leptonic and hadronic components has been proposed as well~\cite{2023arXiv230806994I}. The possible UHECRs in GRB 221009A and the accompanying radiation processes have been also explored~\cite{2023ApJ...944L..34R,2023arXiv230211111W,2023A&A...670L..12D,2022arXiv221012855A}. Therefore, high-energy and especially VHE gamma rays can be a crucial probe for studying leptonic and hadronic dissipation processes inside GRB jets.

Ref.~\cite{doi:10.1126/science.adg9328} have implemented a leptonic scenario, including the electron synchrotron radiations and the self-synchrotron Compton (SSC) scatterings to explain the multi-wavelength EM observations of GRB 221009A. However, the theoretical spectra present apparent discrepancies from the observations at the VHE band, showing harder spectral shapes than the observed values as they indicated. In this work, we implement multiband data analyses. Combining the multi-wavelength observations, we, respectively, study the interactions of prompt emissions with high-energy protons and electrons accelerated in the early afterglow phase by hadronic processes and external inverse Compton (EIC), including the subsequent EM cascade initiated by the secondaries of hadronic processes and the high-energy gamma rays of the EIC process. In addition, the afterglow SSC emission is also involved. We find that the combination of the EM cascade emission and the afterglow SSC emission can well regenerate the multi-wavelength spectra and the temporal light curve.

\section*{Results}\label{sec:results}

\begin{figure*} %[htb!]
	\includegraphics[width=0.99\textwidth]{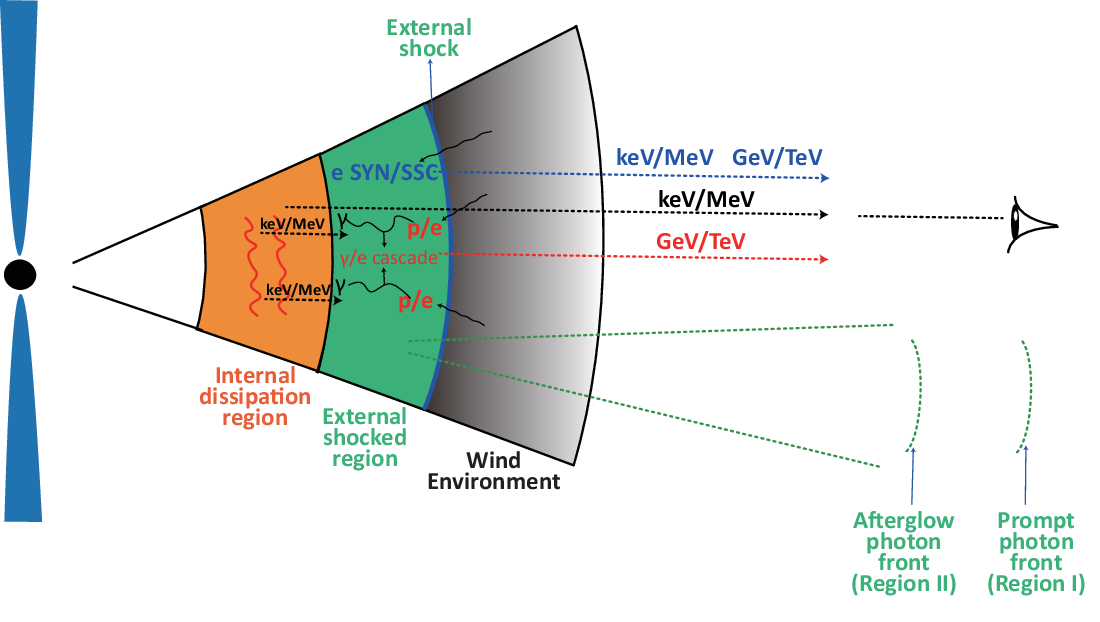} 
	\caption{
		\textbf{Schematic (not to scale) of radiation processes in the external shocked region.} High-energy protons and electrons are accelerated by the external shock. We consider two dissipation regions in the external shocked region. In Region I, the prompt keV/MeV photons catch up with the external shocked region and interact with the accelerated protons and electrons. The secondary gamma rays and electrons produced by the hadronic and EIC processes then initiate the EM cascade to contribute to the observed GeV/TeV emission. In Region II, the classical afterglow emission including the synchrotron and SSC radiations of electrons can contribute to the keV/MeV and GeV/TeV emissions. Region II is behind the Region I. The external forward shock propagates in a stellar wind environment. Note that the jet opening angle is exaggerated for visualization and the true jet opening angle is much smaller.}
	\label{fig:schematic}
\end{figure*}

For GRB 221009A, the GeV and TeV emissions may follow a temporal evolution of afterglow that is characterized by a power-law decay in time \cite{2023arXiv230303855S,doi:10.1126/science.adg9328}, corresponding to the energy dissipation of the external shock caused by the interaction of the GRB jet with the ambient medium. However, MeV emissions exhibit rapid timescale variability \cite{2023arXiv230314172L} and are thought to be produced in the internal dissipation of the GRB jet. One strange thing is the similar rise time ($\sim 220\,\rm s$ after the GBM trigger time $T_0$) of the main bursts of keV/MeV, GeV, and TeV, although the TeV emissions detected by LHAASO have a few seconds delay. The simultaneous arrival of keV/MeV, GeV, and TeV emissions indicates that the prompt keV/MeV emission can catch up with the external dissipation region when the GeV and TeV are generated therein. The prompt photons will overshoot the keV/MeV radiations generated by the external shock itself and can provide a more dominant target photon field at the external dissipation region. We refer to this region as the dissipation \textit{Region I} where at keV/MeV energy band is dominant by the prompt photons. Besides, the afterglow emission could be dominant at another dissipation region, where the seed photons are mainly contributed by the external shock itself and served as the dominant target photon field for the SSC scatterings of accelerated electrons. This region is labeled as the dissipation \textit{Region II}.

As illustrated in Fig.~\ref{fig:schematic}, we suppose that the high-energy protons and electrons can be accelerated by the external shock. In Region I, the prompt keV/MeV photons can catch up with this region and become the dominant target photon field. High-energy and VHE gamma-rays are contributed by the EM cascade initiated by the secondary high-energy photons and electrons of hadronic and EIC processes in the external dissipation region. The $\gamma\gamma$ annihilation for high-energy photons and the synchrotron and IC process for high-energy electrons for the EM cascade are taken into account. In Region II, the target photon field is dominated by the afterglow emission, and the synchrotron and SSC emission of the afterglow model are calculated and will contribute keV/MeV and GeV/TeV observations. The prompt photon front is ahead of the afterglow photon front due to its earlier arrival. The onset time of afterglow emission is set as the same as the time of the brightest prompt gamma-ray peak, namely, $\sim 231\,\rm s$ \cite{2023arXiv230301203A} after Gamma-ray Burst Monitor (GBM) trigger, since the afterglow emission is mainly produced by the most powerful GRB plasma.

%\begin{figure*}
%	\includegraphics[width=0.555\textwidth]{specmodeling.eps}
%	\includegraphics[width=0.555\textwidth]{speccompare.eps}
%	\caption{\textbf{Multi-wavelength spectral modeling}. \emph{Left panel}: The GECAM/Fermi-GBM, Fermi-LAT, and LHAASO observations in four time intervals are presented. The keV/MeV emissions are attributed to the electron synchrotron radiation in the internal shock region, while GeV/TeV emissions are from the hadron-initiated EM cascade in the external shocked region. The Green keV/MeV and GeV observations are derived by Fermi-GBM and Fermi-LAT for a time range $326-650\,\rm s$, but they are scaled to $326-900\,\rm s$ considering a subdominant contribution of $650-900\,\rm s$ \cite{2023arXiv230301203A}. The Fermi-LAT observations in $231-240\,\rm s$ and $248-326\,\rm s$ are treated as the lower limits due to the pile-up effect in $231-236\,\rm s$ and $257-265\,\rm s$. For the time interval $326-650\,\rm s$, a very narrow time-bin under the pile-up effect around $510\,\rm s$ is neglected~\cite{2023arXiv230314172L}, and in order to cross-check, we also implement the spectral analysis of GECAM for this time interval, which is almost same with the Fermi-GBM's. The (EBL corrected) LHAASO data and the SSC theoretical expectation (dotted lines) are taken from Ref.~\cite{doi:10.1126/science.adg9328}. \emph{Right panel}:  comparison of the SSC emission and the hadron-initiated EM cascade in the external shocked region at the TeV energy band.}
%	\label{fig:spec}
%\end{figure*}

\begin{figure*}
	\includegraphics[width=0.500\textwidth]{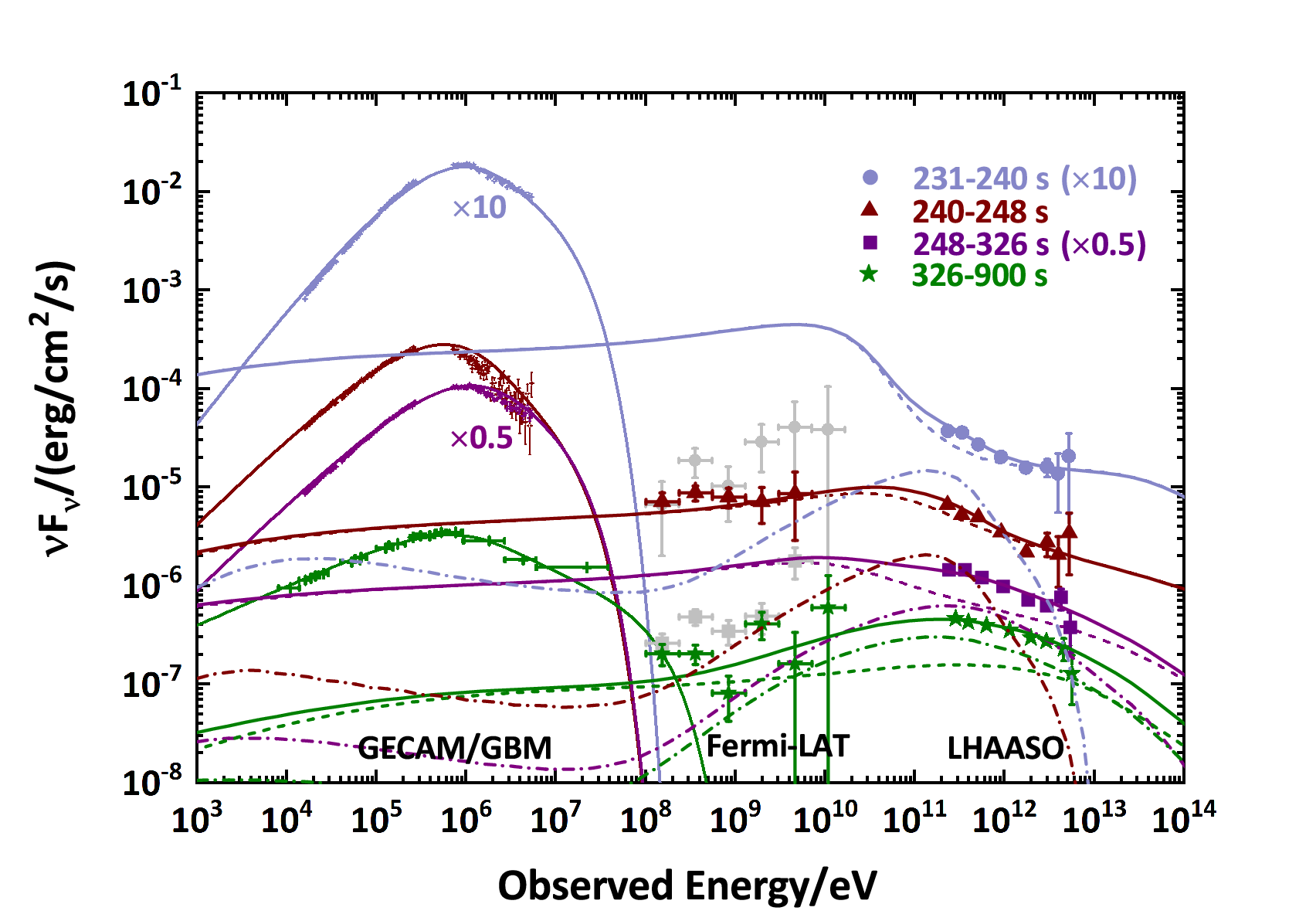}
	\includegraphics[width=0.500\textwidth]{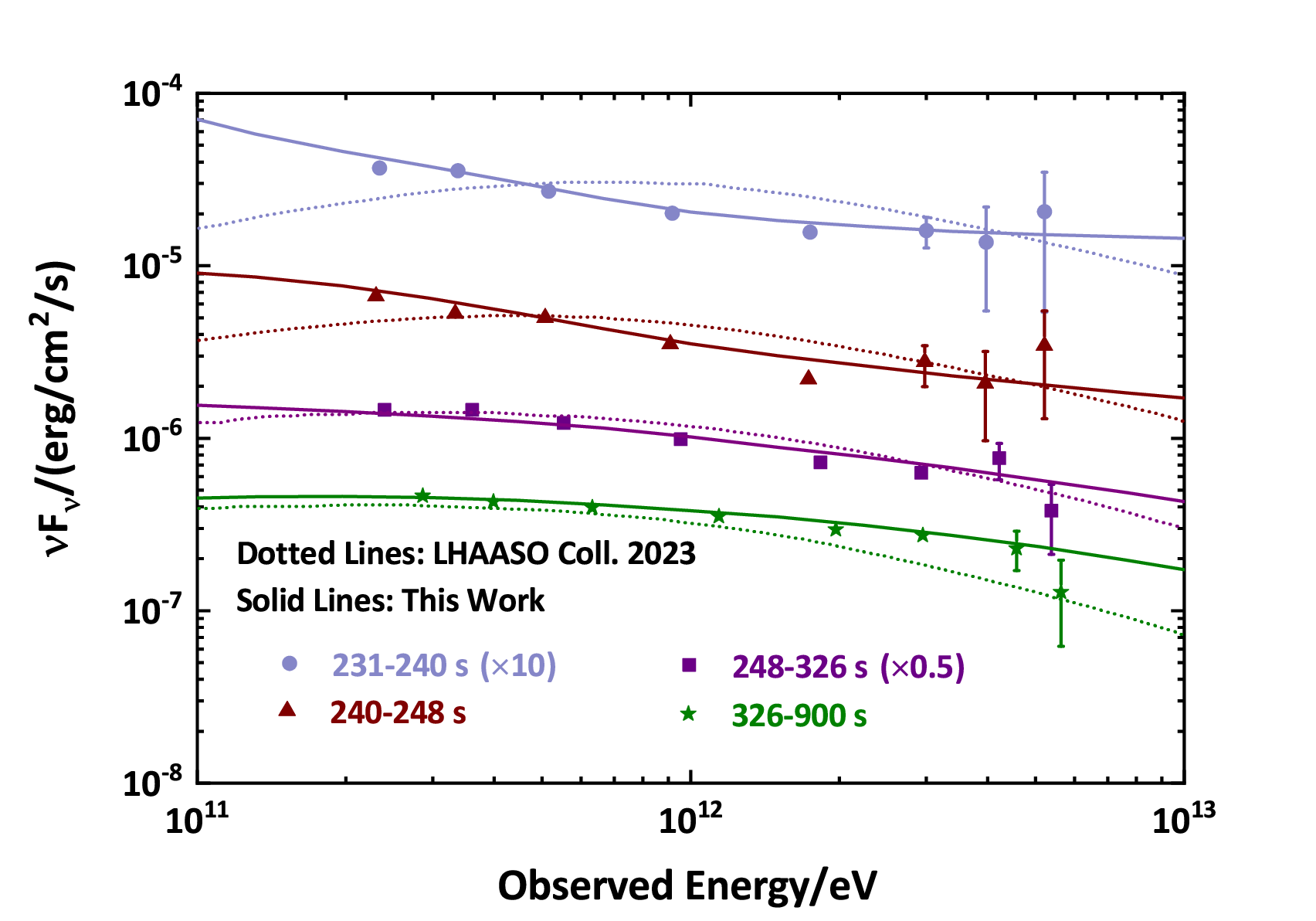}
	\caption{\textbf{Multi-wavelength spectral modeling}. \emph{Left panel}: The GECAM/Fermi-GBM, Fermi-LAT, and LHAASO observations in four time intervals are presented. The keV/MeV emissions are attributed to the electron synchrotron radiation in the internal dissipation region. The hadron-initiated EM cascade is presented by dashed lines, the afterglow emission (including the synchrotron and SSC emissions) by dot-dashed lines, and the sum of both by solid lines. Green keV/MeV and GeV observations are derived by Fermi-GBM and Fermi-LAT for a time range of $326-650\,\rm s$, but they are scaled to $326-900\,\rm s$ considering a subdominant contribution of $650-900\,\rm s$ \cite{2023arXiv230301203A,2023arXiv231010522Z}. The Fermi-LAT observations in $231-240\,\rm s$ and $248-326\,\rm s$ are treated as lower limits due to the pile-up effect in $231-236\,\rm s$ and $257-265\,\rm s$. For the time interval $326-650\,\rm s$, a very narrow time-bin under the pile-up effect around $510\,\rm s$ is neglected~\cite{2023arXiv230314172L}, and in order to cross-check, we also implement the spectral analysis of GECAM for this time interval, which is almost same with the Fermi-GBM's. The (EBL corrected) LHAASO data are taken from Ref.~\cite{doi:10.1126/science.adg9328}. \emph{Right panel}: comparison of the single SSC emission (dotted lines) in Ref.~\cite{doi:10.1126/science.adg9328} and the combination (solid lines) of the hadron-initiated EM cascade and the afterglow emission in the external shocked region at the TeV energy band.}
	\label{fig:spec}
\end{figure*}
In order to derive the multi-wavelength observational data, we analyze the observational data of Gravitational-wave high-energy Electromagnetic Counterpart All-sky Monitor (GECAM) that is free of data saturation or pulse pileup problems during the detection of GRB 221009A \cite{2023arXiv230301203A}, and GBM and Large Area Telescope (LAT) onboard the Fermi satellite. We perform a spectral analysis of GECAM, Fermi-GBM, and Fermi-LAT in three time intervals, that is, $231-240\,\rm s$, $240-248\,\rm s$, and $248-326\,\rm s$ from the GBM trigger, following the same time interval as for LHAASO \cite{doi:10.1126/science.adg9328}. Besides, a spectral analysis in the time interval of $326-650\,\rm s$ is implemented as well for Fermi-GBM and Fermi-LAT, which is scaled to $326-900\,\rm s$ as the same as the time interval for LHAASO. The spectra for different time intervals are presented in Fig.~\ref{fig:spec}. Since the Fermi-LAT data in $231-236\,\rm s$ and $257-265\,\rm s$ is under the pile-up effect, the Fermi-LAT spectra in the time intervals of $231-240\,\rm s$ and $248-326\,\rm s$ are treated as the lower limits, showing as the gray points in Fig.~\ref{fig:spec}.

Moreover, we quantify the temporal correlation between the Fermi-LAT and LHAASO observations, a strong coincidence between the high-energy gamma rays at the Fermi-LAT energy band ($100\,\mathrm{MeV}-\sim 10\,\mathrm{GeV}$) and the VHE gamma rays at the LHAASO Water Cherenkov Detector Array (WCDA) energy band ($\sim 200\,\mathrm{GeV}-7\,\mathrm{TeV}$) are found during the correlation analysis, implying the common origin between both. Therefore, during the spectral modeling, The Fermi-LAT and LHAASO observations have to be simultaneously interpreted.

Combining the observations of GECAM/Fermi-GBM, Fermi-LAT, and LHAASO, we implement multiwavelength spectral modelings for diverse time intervals. The afterglow emission in Region II presents a relatively small contribution to the TeV emission at the beginning and becomes dominant at the late stage. Especially, the relatively flat spectral shape from $100\,\rm MeV$ to multi-TeV at the time interval of $240-248\,\rm s$ can be well regenerated by the EM cascade (in Region I) initiated by the secondaries of hadronic processes. A much more consistent spectral shape with the observations around the TeV energy band than the individual afterglow SSC scenario is derived. At the later stage, the afterglow SSC emission (in Region II) plays a more and more important role. The superposition of the hadron-initiated EM cascade and the afterglow SSC emission can well explain the GeV/TeV emissions from Fermi-LAT and LHAASO observations.

The adopted parameters are listed in Table~\ref{table:parameter}. In our modeling, the parameters for the prompt emission are determined by the keV/MeV observations. The temporal evolution of the afterglow depends on the isotropic kinetic energy $E$ and the wind environment $A$. In Region I, for a typical magnetic energy factor $f_{\rm B, ex}$ and a proton index $s$, the only left free parameter to affect GeV/TeV emissions is the proton kinetic luminosity $L_p$, which will impact the amplitude of the GeV/TeV emission. In our calculations, a large isotropic proton luminosity is involved. However, it is feasible if the jet opening angle for Region I is small. The small jet opening angle has been indicated in Ref.~\cite{doi:10.1126/science.adg9328,2023arXiv230301203A}. Moreover, the cooling rate of accelerated protons is low as shown in Fig.~\ref{fig:timescales}, therefore high-energy protons can be accelerated at a small radius due to the denser gas (more protons can participate in the acceleration) at the small radius for a wind environment. Then they continuously participate in the hadronic interaction and generate the secondaries. For the afterglow emission in Region II, the same isotropic kinetic energy $E$ and the wind environment $A$ is employed. We implement a detailed simulation for the afterglow model as described in Section Methods.

Because the EM cascade is fully developed, the cascade emission presents a universal flat spectral shape. Around the TeV energy band, the descent as the observed energy increases is induced by the absorption of the low-energy keV/MeV emission. As a result, at the early time, a natural relatively flat connection between the GeV and TeV emission and a descent around TeV can be expected. The observational spectral shape can be naturally reproduced by our model. In addition, we calculate the expected accompanying neutrino detection by IceCube based on its effective area at $100\,\mathrm{GeV}- 10\,\mathrm{EeV}$ for a point source at the declination of GRB 221009A ($\delta =19.8^{\circ}$) \cite{2021arXiv210109836I}. The expected neutrino detection number is basically small, consistent with the observational neutrino limits~\cite{2023ApJ...946L..26A}.

\begin{figure*} %[htb!]
	\includegraphics[width=0.99\textwidth]{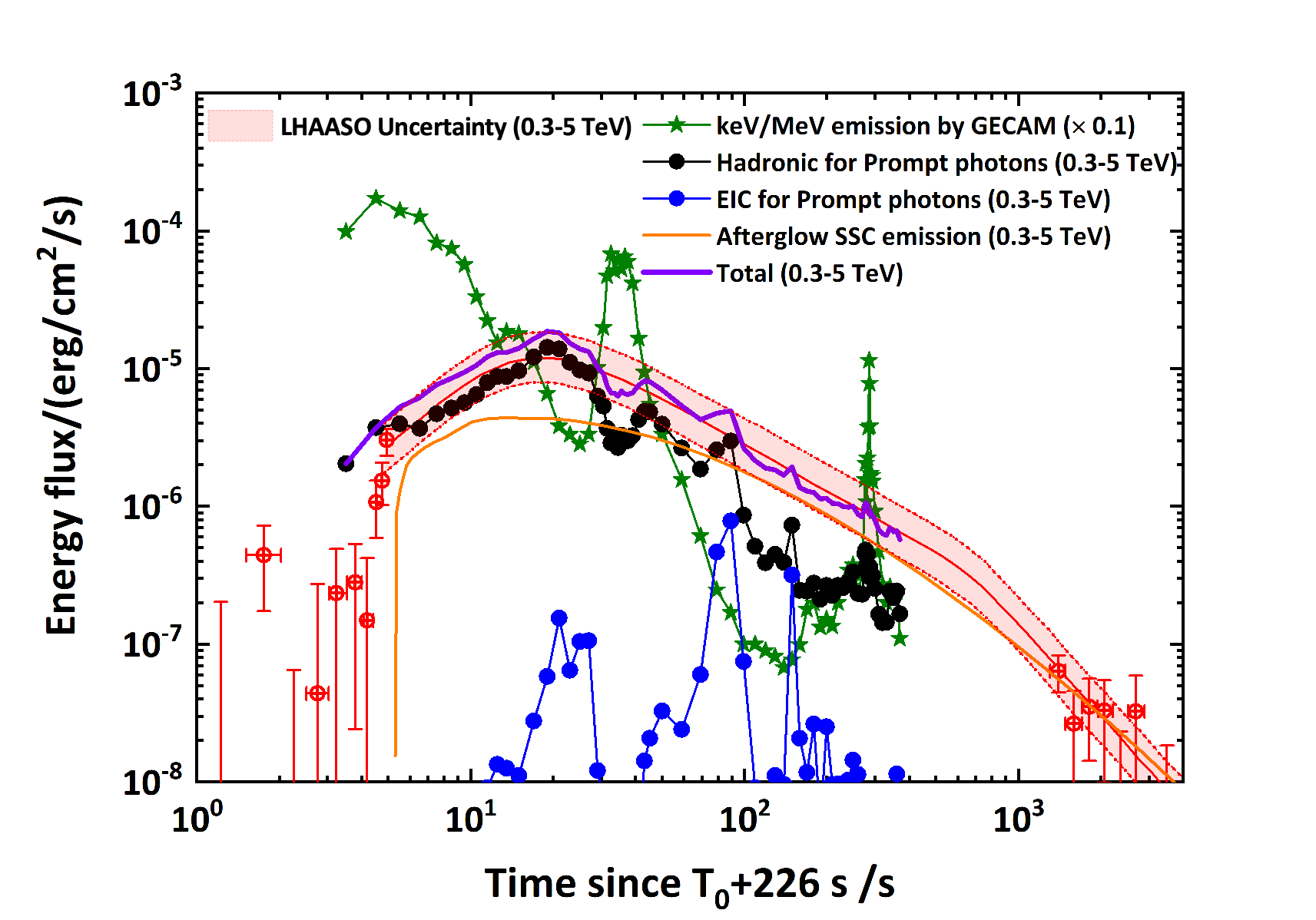} 
	\caption{
		\textbf{TeV light curve for different components.} The red points and shadow region correspond to the uncertainty of LHAASO observations at the energy range of $0.3-5\,\rm TeV$~\cite{doi:10.1126/science.adg9328}, including the data fluctuation and the uncertainties of EBL models. The keV/MeV observations, including the photon indexes, the peak energy, and the energy flux in $1\,\mathrm{keV}-10\,\mathrm{MeV}$, are derived by the GECAM data analyses for diverse time bins. The time bins are selected as small as 1 second for the fast variability period, while larger for the relatively smooth period. The energy flux in $1\,\mathrm{keV}-10\,\mathrm{MeV}$ are presented as the green stars. The black points represent the TeV emission from the hadron-initiated EM cascade, induced by the interactions between the accelerated protons and prompt seed photons. The blue points indicate the TeV emission produced by the accelerated electrons and prompt seed photons through the EIC process in the same dissipation region (Region I) with the hadron-initiated EM cascade process. Both components are calculated based on the keV/MeV seed photons derived by GECAM for the corresponding time bin. The adopted parameters about the temporal evolution of protons and electrons in Region I are $\alpha_1=0.4$, $\alpha_2=-1.2$, $L_{p,c}=10^{55}\,\rm erg/s$, and $L_{e,c}=L_{p,c} m_e/m_p$. In addition, the afterglow SSC emission (Region II) produced by the accelerated electrons is taken into account and presented as the orange solid line. the onset time of the afterglow SSC emission is set as the same as the time of the brightest prompt gamma-ray peak, says, $231\,\rm s$ after GBM trigger. For the afterglow SSC emission, the adopted parameters are listed in Table~\ref{table:parameter}.}
	\label{fig:light curve}
\end{figure*}

For further study of the origin of the TeV emission, we reproduce the TeV light curve in Fig.~\ref{fig:light curve}. We analyze the GECAM data for small time bins and derive the prompt keV/MeV photon distributions, which are used as the target-photon field in Region I for the hadron-initiated EM cascade and the EIC process. As we can see in Fig.~\ref{fig:light curve}, the TeV emission from the hadron-initiated EM cascade is relatively insensitive to the prompt seed photons with significant variabilities, since the increase in hadronic interaction efficiency for a high prompt seed photon flux is canceled by the photon-photon annihilation. However, such a TeV variability could be large when the internal absorption inside the GRB jet becomes weak, e.g. a weaker keV/MeV emission and a larger dissipation radius at the later stage. On the whole, the hadron-initiated EM cascade presents a smooth light curve at the early stage and clear temporal variability at the late stage. Therefore, the hadron-initiated EM cascade cannot be responsible for the entire TeV emission observed by LHAASO. For our model, the afterglow emission from Region II will be dominant in the TeV energy band at the late stage and presents a smooth TeV light curve. The superposition of two components from Region I and Region II can generate a slight variability in the TeV light curve, which seems consistent with the LHAASO observations that indicate the persistent small fluctuation of the TeV emission.

The individual afterglow emission basically presents a hard spectral shape and is not consistent with the observational GeV/TeV emissions, which show a quite flat connection from GeV to TeV at the early stage. The Fermi-LAT and LHAASO observations present a strong correlation based on our analyses, which have to be explained simultaneously. Second, even at the TeV energy band, clear deviations of theoretical spectra from the observations, especially for the time intervals of $231-240\,\rm s$ and $240-248\,\rm s$ can be seen from Fig.~\ref{fig:spec} since the SSC spectral bump is relatively hard. In contrast, the individual hadron-initiated EM cascade can present a natural flat spectral shape from GeV to TeV and a smooth TeV light curve at the early stage, whereas a too-large TeV variability at the late stage is not consistent with the relatively smooth observations of the TeV light curve. In summary, our model combined the hadron-initiated EM cascade with the afterglow emission can well regenerate the multi-band spectra for diverse time intervals and consistent light curves.

\begin{table*}
 \centering 
    \footnotesize
	\caption{The adopted parameters in spectral modeling for the time intervals of $231-240\,\rm s$, $240-248\,\rm s$, $248-326\,\rm s$, and $326-900\,\rm s$.}
    \vspace{0.3cm}
	\begin{tabular}{l@{\hskip 0.1in}l@{\hskip 0.1in}l@{\hskip 0.1in}l}
 \hline
  & Descriptions	&	Symbols	&	Values	\\
 \hline
Observations	&	Redshift	&	z	&	0.15 \\
	&	Low energy photon index	&	$\alpha_\gamma$	&	[-0.91, -1.19, -1.2,-1.59]	\\
	&	High energy photon index	&	$\beta_\gamma$	&	[-2.396, -2.7, -2.38, -2.43]	\\
	&	Peak energy	&	$\varepsilon_{\gamma,p}$	&	[688, 560, 966, 698]\,MeV	\\
 	&	Luminosity	&	$L_{\gamma}$$^a$	&	[35, 5.5, 4, 0.2]$\times 10^{52}\,\rm erg \,s^{-1}$	\\
 \hline
Prompt	&	Variability timescale	&	$\delta t$	&	$0.082\,\rm s$	\\
        &	Internal bulk Lorentz factor	&	$\Gamma_{\rm in}$$^b$	&	520	\\
	&	Internal dissipation radius	&	$R_{\rm in}$	&	$2\Gamma_{\rm in}^2 c \delta t$	\\
	&	Low energy electron index	&	$\alpha_e$	&	[-0.3, -1.2, -1.2, -2.2]	\\
	&	High-energy electron index	&	$\beta_e$	&	[-4.29, -4.9, -4.26, -3.86]	\\
	&	Electron break energy	&	$\gamma_{e,b}$	&	[1001, 1375, 1108, 1420]	\\
        &	Magnetic energy factor	&	$f_{\rm B, in}$	&	30	\\
        &	Electron energy factor	&	$f_e$$^c$	&	1	\\
\hline
Afterglow &	Proton index	&	$s$	&	-2	\\
(Region I)	&	Proton luminosity	&	$L_p$	&	[3.67, 5.8, 1.6, 0.7]$\times 10^{54}\,\rm erg \,s^{-1}$	\\
	&	Magnetic energy factor	&	$f_{\rm B, ex}$   &	0.1	\\
	&	Isotropic kinetic energy	&	$E$	&	$10^{55} \,\rm erg$	\\
	&	Wind parameter	&	$A$	&	$10^{35.5} \,\rm cm^{-1}$	\\
\hline
Afterglow &	Isotropic kinetic energy	&	$E$	&	$10^{55} \,\rm erg$	\\
(Region II)	&	Wind parameter	&	$A$	&	$10^{35.5} \,\rm cm^{-1}$	\\
        &	Initial bulk Lorentz factor	&	$\Gamma_0$	&	$260$	\\
	&	Electron injection index	&	$p$   &	2.42	\\
	&	Jet opening angle	&	$\theta_j$	&	$10^{-1.55}$	\\
	&	Electron energy fraction	&	$\varepsilon_e$	&	$10^{-1.64}$	\\
         &	Magnetic energy fraction	&	$\varepsilon_B$	&	$10^{-5.97}$	\\
         &	Non-thermal electron fraction	&	$\xi_e$	&	$10^{-0.59}$	\\
\hline
Outputs	&	Expected neutrino number	&	$N_\nu$$^d$	&	[0.33,0.02,0.28, 0.11]	\\
 \hline
	\end{tabular}
    
    $^a$ in 1\,keV-10\,MeV;
    $^b$ $\Gamma_{\rm in}$ is roughly adopted as double the bulk Lorentz factor at the deceleration site;
    $^c$ $f_e=L_e/L_{\gamma}$;
    $^d$ in 100\,GeV-10\,EeV.
	\label{table:parameter}
\end{table*}

%\begin{figure*} %[htb!]
%	\includegraphics[width=0.99\textwidth]{Ryde-triangle.pdf} 
%	\caption{
%		\textbf{The best parameters for the afterglow SSC emission.} The same isotropic kinetic energy $E=10^{55}\,\rm erg$, initial bulk Lorentz factor $\Gamma_0=260$, and wind parameter $A=10^{35.5}\,\rm cm^{-1}$ are fixed. The other parameters are determined by the MCMC simulation of the afterglow model and presented in the figure. Here, $p$ is the electron injection index, $\theta_j$ is the jet opening angle, $\varepsilon_e$ is the energy fraction carried by electrons, $\varepsilon_B$ is the energy fraction carried by the magnetic field, and $\xi_e$ is the fraction of accelerated non-thermal electrons to the total electrons.
%  }
%	\label{fig:rydetriangle}
%\end{figure*}

\section*{Conclusions}
We implement the spectral analyses for GRB 221009A based on the observations of GECAM, Fermi-GBM, and Fermi-LAT. Based on the multi-wavelength data, the hadron-initiated EM cascade and EIC process with the prompt seed photons, and the afterglow emission are implemented. A much more consistent spectral shape for diverse time intervals can be regenerated. In the future, when the data of LHAASO larger air shower kilometer square area (KM2A) that can capture $>10 \,\rm TeV$ photons is released, it can be a further check for the origin of VHE gamma-rays. Our model also predicts the possible slight variability in the TeV light curve, so we encourage the search of the TeV emission variability by LHAASO, which can determine the contribution extent of radiation processes with the prompt seed photons and the afterglow SSC emission.

\section*{Methods}\label{sec:methods}

\subsection*{GECAM Data}
% GECAM Data comes from GECAM-C onboard SATech-01 satellite. Which detector used? Time range, energy range.  Note that GECAM-C suffers no pulse pileup effect. The dead time for the LG channel of GECAM-C detectors is also accurate. Details of GECAM-C data and analysis could be found in An et al. (submitted).
GECAM Data comes from GECAM-C onboard SATech-01 satellite. In our data analysis, GRD01(LaBr3) was used, which has been well-calibrated \cite{zhang2023cross}. The available energy range for High Gain is 15-300 keV and 700-5500 keV for Low Gain. We have performed the spectrum fitting in time intervals of 231-240 s, 240-248 s, 248-326 s, and 326-650 s with the classic Band function. For the TeV light curve calculation, we also analyze the GECAM data for small time bins, which are as small as 1 second for the fast variability period and larger for the relatively smooth period with the best distribution function including the Band function, exponential-cut-off power-law, and the single power-law. Notably, the GECAM-C data did not saturate throughout the whole burst, and the dead time for the Low Gain channel of GECAM-C detectors is also accurate. It should be noted that the high gain dead time recording is somewhat problematic during the brightness of the burst, but this does not affect the spectral profiles. Therefore, a constant factor needs to be liberalized to correct this effect when doing energy spectrum analysis. Details of GECAM-C data analysis can be found in Ref.~\cite{2023arXiv230301203A}.

\begin{table}
 \centering 
    \footnotesize
	\caption{Time averaged spectrum fitting results by GECAM for four time intervals.}
    \vspace{0.3cm}
	\begin{tabular}{l@{\hskip 0.1in}l@{\hskip 0.1in}l@{\hskip 0.1in}l@{\hskip 0.1in}l}
 \hline
 Time interval	&	$\alpha$	&	$\beta$	&	$E_{\,\rm cut}$ &	Flux	\\
  (s)	&			&			&	(keV) &	($\rm erg\ cm^{-2}\ s^{-1}$)	\\
 \hline
231-240& $-0.92_{-0.01}^{+0.01}$ & $-2.40_{-0.01}^{+0.01}$ & $688.56_{-27.34}^{+30.32}$ & $3.66_{-0.09}^{+0.08}  \times 10^{-3} $ \\
240-248	& $-1.19_{-0.01}^{+0.01}$ & $-2.72_{-0.03}^{+0.03}$ & $559.62_{-24.17}^{+24.53}$ & $7.18_{-0.23}^{+0.27}  \times 10^{-4} $ \\
248-326	& $-1.15_{-0.01}^{+0.01}$ & $-2.38_{-0.01}^{+0.01}$ & $966.01_{-25.46}^{+30.87}$ & $4.37_{-0.07}^{+0.06}  \times 10^{-4} $ \\
326-650	& $-1.57_{-0.01}^{+0.01}$ & $-2.65_{-0.04}^{+0.05}$ & $634.47_{-20.66}^{+25.59}$ & $4.67_{-0.18}^{+0.14}  \times 10^{-5} $ \\
 \hline
	\end{tabular}
	\label{table:gecam}
 \footnotesize{{\leftline {Flux is calculated in energy range $1\,\mathrm{keV}-10\,\mathrm{MeV}$.}}}\\
\end{table}

\subsection*{Fermi-LAT Data}
Within 12\textdegree of the reported LAT position, R.A.=288.282\textdegree, decl.=19.495\textdegree(J2000), the Pass 8 transient events are used in the energy range of 100 MeV and 300 GeV \cite{GCNlatposition}.
Events with zenith angles $>$100\textdegree are excluded to reduce the contribution of Earth-limb gamma-rays. The instrument
response function “P8R3\_TRANSIENT020\_V3” is used. 

We perform a spectral analysis on the prompt emission in four time intervals, such as 231-240\,s, 240-248\,s, 248-326\,s, and 326-650\,s from the GBM trigger. Note that, events in the time bins 231-236s and 257-265s are excluded \cite{2022GCN.32760....1O}. In every interval, we perform the standard unbinned likelihood analysis. The photon model consists of two sources, a GRB modeled as a point source and a diffuse source with an energy distribution in \textit{iso\_P8R3\_TRANSIENT020\_V3\_v1.txt}. The resultant SEDs are shown in Fig.~\ref{fig:spec}.

\subsection*{Coincidence of Fermi-LAT and LHAASO Data}
In order to quantify the temporal correlation between the Fermi-LAT and LHAASO observations for GRB 221009A, we divided the time range 231-400s into 61 time bins, in which the measured energy fluxes both by the Fermi-LAT and LHAASO are derived. Time bins among 231-236s and 257-265s are excluded.  Fermi-LAT fluxes ($f_{\rm Fermi}$) are shown in Table~\ref{table:flux}. The LHAASO fluxes ($f_{\rm LHAASO}$) in the same time bins are derived from Ref.~\cite{doi:10.1126/science.adg9328}, which can be found in Table~\ref{table:flux}. 

Then the GeV-TeV correlation between the Fermi-LAT energy flux ($f_{\rm F}$) and LHAASO energy flux ($f_{\rm L}$) is tested in all time bins, i.e., ${\rm N_{bin}} = 61$. Assuming $f_{\rm F}(T_i)$ and $f_{\rm L}(T_i)$
are the Fermi-LAT flux and LHAASO flux at the time of $T_i$, the linear equation can be represented as:
\begin{equation}
f_{\rm L}(T_i) = A + B \times f_{\rm F}(T_i)
\end{equation}
According to Pearson's correlation coefficient $R$ can be represented:
\begin{equation}
R =  \frac{\sum_{1}^{\rm N_{bin}}(f_{\rm F}(T_i)-\bar f_{\rm F})(f_{\rm L}(T_i)-\bar f_{\rm L})}
{\sqrt{\sum_{1}^{\rm N_{bin}}(f_{\rm F}(T_i)-\bar f_{\rm F})^2}  \sqrt{\sum_{1}^{\rm N_{bin}}(f_{\rm L}(T_i)-\bar f_{\rm L})^2}}
\end{equation}
where the $\bar f_{\rm F}$ and $\bar f_{\rm G}$ represent respectively the average fluxes of the Fermi-LAT and the LHAASO in the chosen time interval. We also calculate the $p$ value of the null hypothesis using the software of \textit{Origin}, which can be described as the confidence level of $1-p$ for the GeV-TeV correlation \cite{correlationmethod}. In this analysis, the resultant $R$ is 0.87, which indicates a strong positive correlation. Moreover, the $p$ is less than $10^{-4}$, which is for such a strong correlation in deep.

\begin{table}
 \centering 
    \footnotesize
	\caption{Observed energy flux by Fermi-LAT and LHAASO in 61 time intervals.}
    \vspace{0.3cm}
	\begin{tabular}{l@{\hskip 0.1in}l@{\hskip 0.1in}l@{\hskip 0.1in}l}
 \hline
 $T_{\rm Start}$	&	$T_{\rm End}$	&	$f_{\rm Fermi}$	&	$f_{\rm LHAASO}$	\\
  (s)	&	(s)		&	($10^{-6}\,\rm erg\ cm^{-2}\ s^{-1}$)		&	($10^{-6}\,\rm erg\ cm^{-2}\ s^{-1}$)	\\
 \hline
236	&	238	&	17.63	$\pm$	7.73	&	8.49	$\pm$	0.67	\\
238	&	240	&	39.18	$\pm$	11.79	&	10.53	$\pm$	0.67	\\
240	&	242	&	26.29	$\pm$	9.30	&	12.01	$\pm$	0.95	\\
242	&	244	&	27.31	$\pm$	6.75	&	11.81	$\pm$	1.12	\\
244	&	246	&	53.70	$\pm$	13.39	&	12.21	$\pm$	0.97	\\
246	&	248	&	54.36	$\pm$	12.56	&	11.81	$\pm$	1.80	\\
248	&	250	&	61.78	$\pm$	13.86	&	10.69	$\pm$	1.63	\\
250	&	252	&	33.20	$\pm$	9.67	&	10.17	$\pm$	0.81	\\
252	&	254	&	25.00	$\pm$	8.83	&	9.51	$\pm$	1.32	\\
254	&	256	&	23.59	$\pm$	9.14	&	9.05	$\pm$	1.25	\\
266	&	268	&	9.24	$\pm$	4.65	&	6.39	$\pm$	0.41	\\
268	&	270	&	3.48	$\pm$	1.67	&	6.49	$\pm$	0.52	\\
270	&	272	&	7.81	$\pm$	3.55	&	6.28	$\pm$	0.87	\\
272	&	274	&	13.51	$\pm$	4.42	&	6.07	$\pm$	0.67	\\
274	&	276	&	13.17	$\pm$	5.20	&	5.88	$\pm$	0.56	\\
276	&	278	&	13.02	$\pm$	5.47	&	5.68	$\pm$	0.45	\\
278	&	280	&	8.85	$\pm$	4.52	&	5.50	$\pm$	0.60	\\
280	&	282	&	11.24	$\pm$	4.04	&	5.23	$\pm$	0.65	\\
282	&	284	&	7.96	$\pm$	2.96	&	4.98	$\pm$	0.40	\\
284	&	286	&	4.13	$\pm$	2.97	&	4.74	$\pm$	0.38	\\
286	&	288	&	10.28	$\pm$	4.80	&	4.58	$\pm$	0.29	\\
288	&	290	&	5.64	$\pm$	2.92	&	4.43	$\pm$	0.49	\\
290	&	292	&	5.85	$\pm$	3.06	&	4.29	$\pm$	0.47	\\
292	&	294	&	6.82	$\pm$	2.55	&	4.15	$\pm$	0.39	\\
294	&	296	&	2.49	$\pm$	1.68	&	4.01	$\pm$	0.19	\\
296	&	298	&	5.56	$\pm$	3.17	&	3.88	$\pm$	0.25	\\
298	&	300	&	5.17	$\pm$	3.14	&	3.75	$\pm$	0.24	\\
300	&	302	&	4.84	$\pm$	3.01	&	3.63	$\pm$	0.29	\\
302	&	304	&	8.95	$\pm$	3.94	&	3.69	$\pm$	0.35	\\
304	&	306	&	4.15	$\pm$	2.72	&	3.63	$\pm$	0.50	\\
306	&	308	&	10.03	$\pm$	4.34	&	3.57	$\pm$	0.49	\\
308	&	310	&	7.89	$\pm$	3.31	&	3.45	$\pm$	0.38	\\
310	&	312	&	3.06	$\pm$	2.02	&	3.34	$\pm$	0.32	\\
312	&	314	&	6.11	$\pm$	3.39	&	3.23	$\pm$	0.26	\\
314	&	316	&	4.41	$\pm$	2.28	&	3.18	$\pm$	0.30	\\
316	&	318	&	2.56	$\pm$	2.21	&	3.08	$\pm$	0.29	\\
318	&	322	&	1.38	$\pm$	0.89	&	2.88	$\pm$	0.18	\\
322	&	326	&	4.38	$\pm$	2.44	&	2.78	$\pm$	0.13	\\
326	&	328	&	10.79	$\pm$	6.36	&	2.78	$\pm$	0.18	\\
328	&	330	&	3.71	$\pm$	1.93	&	2.78	$\pm$	0.18	\\
330	&	332	&	9.84	$\pm$	6.21	&	2.69	$\pm$	0.17	\\
332	&	334	&	2.85	$\pm$	1.75	&	2.60	$\pm$	0.21	\\
334	&	336	&	7.74	$\pm$	5.63	&	2.56	$\pm$	0.20	\\
336	&	338	&	4.50	$\pm$	2.05	&	2.48	$\pm$	0.20	\\
338	&	340	&	6.82	$\pm$	5.12	&	2.78	$\pm$	0.30	\\
340	&	342	&	2.90	$\pm$	1.80	&	2.40	$\pm$	0.23	\\
342	&	344	&	2.47	$\pm$	2.21	&	2.40	$\pm$	0.25	\\
344	&	346	&	7.03	$\pm$	3.34	&	2.28	$\pm$	0.24	\\
346	&	350	&	2.45	$\pm$	2.22	&	2.21	$\pm$	0.21	\\
350	&	354	&	3.36	$\pm$	2.58	&	2.10	$\pm$	0.13	\\
354	&	356	&	5.88	$\pm$	4.86	&	1.72	$\pm$	0.18	\\
356	&	358	&	3.00	$\pm$	2.94	&	1.72	$\pm$	0.14	\\
358	&	366	&	1.18	$\pm$	1.08	&	1.72	$\pm$	0.14	\\
366	&	368	&	1.80	$\pm$	1.30	&	1.93	$\pm$	0.24	\\
368	&	372	&	2.09	$\pm$	2.00	&	1.81	$\pm$	0.22	\\
372	&	374	&	2.10	$\pm$	1.70	&	1.72	$\pm$	0.21	\\
374	&	376	&	3.80	$\pm$	2.02	&	1.64	$\pm$	0.18	\\
376	&	386	&	1.47	$\pm$	1.10	&	1.30	$\pm$	0.10	\\
386	&	390	&	0.87	$\pm$	0.63	&	1.43	$\pm$	0.18	\\
390	&	394	&	2.10	$\pm$	1.58	&	1.48	$\pm$	0.13	\\
394	&	400	&	1.70	$\pm$	1.35	&	1.28	$\pm$	0.10	\\
 \hline
	\end{tabular}
	\label{table:flux}
\end{table}

%\FloatBarrier

\subsection*{Modeling of multi-wavelength spectra}

In region I, we consider an isotropically expanding shell with the initial bulk Lorentz factor $\Gamma_0$ from the central engine, interacting with a circum-burst wind-like external medium. The observed keV/MeV radiations are generated by internal dissipation due to their fast timescale variability. The GeV/TeV emissions are produced by the EM cascade of secondary gamma-rays and electrons of hadronic processes between the accelerated protons by the external forward shock and the catch-up keV/MeV radiation field.

The spectra of keV/MeV radiations can be basically depicted by a broken-power-law distribution, i.e., $dn_\gamma/d\varepsilon_\gamma=A_\gamma(\varepsilon_\gamma/\varepsilon_{\gamma,p})^{q_\gamma}$ with a peak energy $\varepsilon_{\gamma,p}$, a low-energy index $q_\gamma=\alpha_\gamma$ for $\varepsilon_\gamma<\varepsilon_{\gamma,p}$ and a high-energy photon index $q_\gamma=\beta_\gamma$ for $\varepsilon_\gamma>\varepsilon_{\gamma,p}$. The normalized coefficient is $A_\gamma = {\Gamma^2U_\gamma }/\left[ {\int_{{\varepsilon _{\gamma,\min }}}^{{\varepsilon _{\gamma,\max }}} {(\varepsilon_\gamma/\varepsilon_{\gamma,p})^{q}{\varepsilon _\gamma }d{\varepsilon _\gamma }} } \right]$, where $U_\gamma =L_\gamma /(4 \pi R^2 \Gamma^2 c)$ is the photon energy density in the comoving frame, and $L_\gamma$ is the luminosity integrated from $\varepsilon _{\gamma,\min }$ to $\varepsilon _{\gamma,\max}$, which are fixed to be $1\,\rm keV$ and $10\,\rm MeV$ for calibration respectively. Here, we employ the synchrotron emission of non-thermal electrons to explain the observed keV/MeV emissions, although some other radiation mechanism is suggested as well, e.g., the photospheric emission~\cite{2011ApJ...732...49P,2013ApJ...765..103L,2013MNRAS.428.2430L}, the Comptonized quasi-thermal emission from the photosphere~\cite{2005ApJ...628..847R,2014ApJ...785..112D} (see a review, e.g., \cite{2014IJMPD..2330002Z}). For the phenomenological spectral fittings, we simply introduce accelerated non-thermal electrons with the broken-power-law distribution, i.e., $dn_e/d\gamma_e=A_e(\gamma_e/\gamma_{e,b})^{q_e}$with a break electron Lorentz factor $\gamma_{e,b}$, a low-energy index $q_e=\alpha_e$ for $\gamma_e<\gamma_{e,b}$ and a high-energy electron index $q_e=\beta_e$ for $\gamma_e>\gamma_{e,b}$, although the low-energy electron index may deviate from the expectation of the standard synchrotron fast cooling, which may need to invoke the possible evolutional magnetic field in the post-shock region~\cite{2014NatPh..10..351U,2021Galax...9...68W}. In addition, we also calculated the self-synchrotron absorption (SSA) frequency $\varepsilon_{\gamma,\rm SSA}$ \cite{1979rpa..book.....R}. Below the SSA energy $\varepsilon_{\gamma,\rm SSA}$, the photon distribution is introduced as $dn_\gamma/d\varepsilon_\gamma \propto \varepsilon_\gamma/\varepsilon_{\gamma,\rm SSA}$ \cite{10.1046/j.1365-8711.2003.06602.x}.

Motivated by the limitation on the prompt emission phase~\cite{2023arXiv230211111W,doi:10.1126/science.adg9328}, a highly magnetized jet is favored. Thus a large magnetic energy factor (defined as the ratio of magnetic luminosity to keV/MeV photon luminosity) $f_{\rm B,in}=L_{\rm B,in}/L_\gamma=30$ is involved. Such a larger magnetic field will induce the SSC flux from the internal shocked region to be greatly suppressed with a negligible contribution to the observed GeV/TeV emissions. The internal dissipation radius can be evaluated by $R_{\rm in}=2\Gamma_0 c \delta t$ based on its variability timescale of the prompt emission phase, which is found to be $\delta t\sim 0.082 \,\rm s$ \cite{2023ApJ...943L...2L}.

Since the GeV/TeV emissions follow the temporal evolution of the afterglow phase and the leptonic scenario presents theoretical spectra deviating from observations. Here, we propose the GeV/TeV emissions are generated by the hadronic processes and the subsequent EM cascade initiated by their secondaries in the external shocked region. The primary protons are assumed to be accelerated by the external forward shock to a power-law with a highest-energy exponential cutoff, i.e., $d{n_p}/d\gamma _p = {A_p}{\gamma _p}^s\exp{(-\gamma_{p}/\gamma_{p,\max})}$, $\gamma_{p,\max}$ is determined by the balance between the acceleration timescale and the cooling timescale (or the dynamical timescale), namely, ${t_{\rm acc}} = \min \{ {t_{\rm cooling}},{t_{\rm dyn}}\} $. The dynamical timescale in the comoving frame of the post-forward-shock region is $t_{\rm dyn}\simeq R_{\rm ex}/\Gamma c$.  The comoving acceleration timescale is ${{t}_{\rm acc}} \simeq \eta {{\gamma }_p}{m_p}c/eB$ in the magnetic field strength $B$, where $e$ is the electron charge and $\eta(\ge 1)$ is the Bohm factor which indicates the deviation from the acceleration in the Bohm diffusion. Here, the Bohm diffusion ($\eta=1$), which means the Larmor radius equals the correlation length of the magnetic field, is employed. For the cooling processes of protons, the synchrotron radiation, the photomeson production process, and the BH process are considered. The comoving synchrotron cooling timescale for the relativistic proton can be written as $t_{\rm syn} = \frac{9(\gamma_p-1)}{4{\gamma_p} ^2}\frac{{m_p}^3 c^5}{e^4B^2}$. The photomeson production and BH timescales are calculated by integrating their productions following the semi-analytical treatment provided by Ref.~\cite{2008PhRvD..78c4013K}. The magnetic field in the external shocked region is introduced by a magnetic energy factor as same as for the prompt emission phase, namely, $f_{\rm B,ex}=L_{\rm B,ex}/L_\gamma$, and then it can be achieved by $B_{\rm ex}=\sqrt{2f_{\rm B,ex} L_\gamma/\Gamma^2R_{\rm ex}^2 c}$. The primary electrons can also be accelerated by the external forward shock, but their radiations are treated as the subdominant contribution.

The afterglow light curve modeling suggests a highly-relativistic external forward shock sweeping a wind-like medium \cite{2023ApJ...947...53R,Laskar_2023}. We adopt the same assumption that the TeV afterglow onset time is $T^*=T_0+226$ as in Ref.~\cite{doi:10.1126/science.adg9328}, where $T_0$ is the Fermi-GBM trigger time. The TeV peak time ($t_{\rm peak}\sim 18\,\rm s$ after $T^*$) corresponds to the deceleration time of the GRB jet when the jet energy is transferred to the shocked external medium. Therefore, the bulk Lorentz factor at the deceleration site can be evaluated by $\Gamma_{\rm d}  \simeq 260{(1 + z)^{1/4}}E_{55}^{1/4}A_{35.5}^{ - 1/4}{\left( {{t_{\rm peak}}/{{18\,\rm s}}} \right)^{ - 1/4}}$,
and the deceleration radius is
$R_{\rm d} \simeq 6 \times {10^{16}}{(1 + z)^{ - 1/2}}E_{55}^{1/2}A_{35.5}^{ - 1/2}{\left( {{t_{\rm peak}}/{{18\,\rm s}}} \right)^{ 1/2}}{\,\rm{cm}}$, where $E$ is the isotropic kinetic energy, $A$ is the wind parameter (the number density of the external medium is $n(r)=A r^{-2}\,\rm{cm^{-3}}$ with a radius $r$ from the central engine), and the conventional expression $Q_x=Q/10^{x}$ is used in cgs units. The initial bulk Lorentz factor is assumed as $\Gamma_0=\Gamma_{\rm d}$, and before the deceleration radius, the external forward shock is in the coasting stage. Therefore, the evolution of the bulk Lorentz factor and the dissipation radius can be summarized as
\begin{equation}
\Gamma  = \left\{ \begin{array}{ll}
\Gamma_0=260{(1 + z)^{1/4}}E_{55}^{1/4}A_{35.5}^{ - 1/4}   & t < 18{\rm{s}}\\
260{(1 + z)^{1/4}}E_{55}^{1/4}A_{35.5}^{ - 1/4}{\left( {{t}/{{18\,\rm s}}} \right)^{ - 1/4}}            & t \ge  18{\rm{s}}
\end{array} \right.
\label{Eqgamma}
\end{equation}

\begin{equation}
R_{\rm ex}  = \left\{ \begin{array}{llll}
6 \times {10^{16}}{(1 + z)^{ - 1/2}}E_{55}^{1/2}A_{35.5}^{ - 1/2}-2\Gamma_{0}^2 c  (18-t)&{\,\rm{cm}} \\  & t < 18{\rm{s}}\\
6 \times {10^{16}}{(1 + z)^{ - 1/2}}E_{55}^{1/2}A_{35.5}^{ - 1/2}{\left( {{t}/{{18\,\rm s}}} \right)^{ 1/2}}&{\,\rm{cm}}   \\        & t \ge 18{\rm{s}}
\end{array} \right.
\label{Eqradius}
\end{equation}
with the time $t=T-T^*$.

\begin{figure} %[htb!]
	\includegraphics[width=\columnwidth]{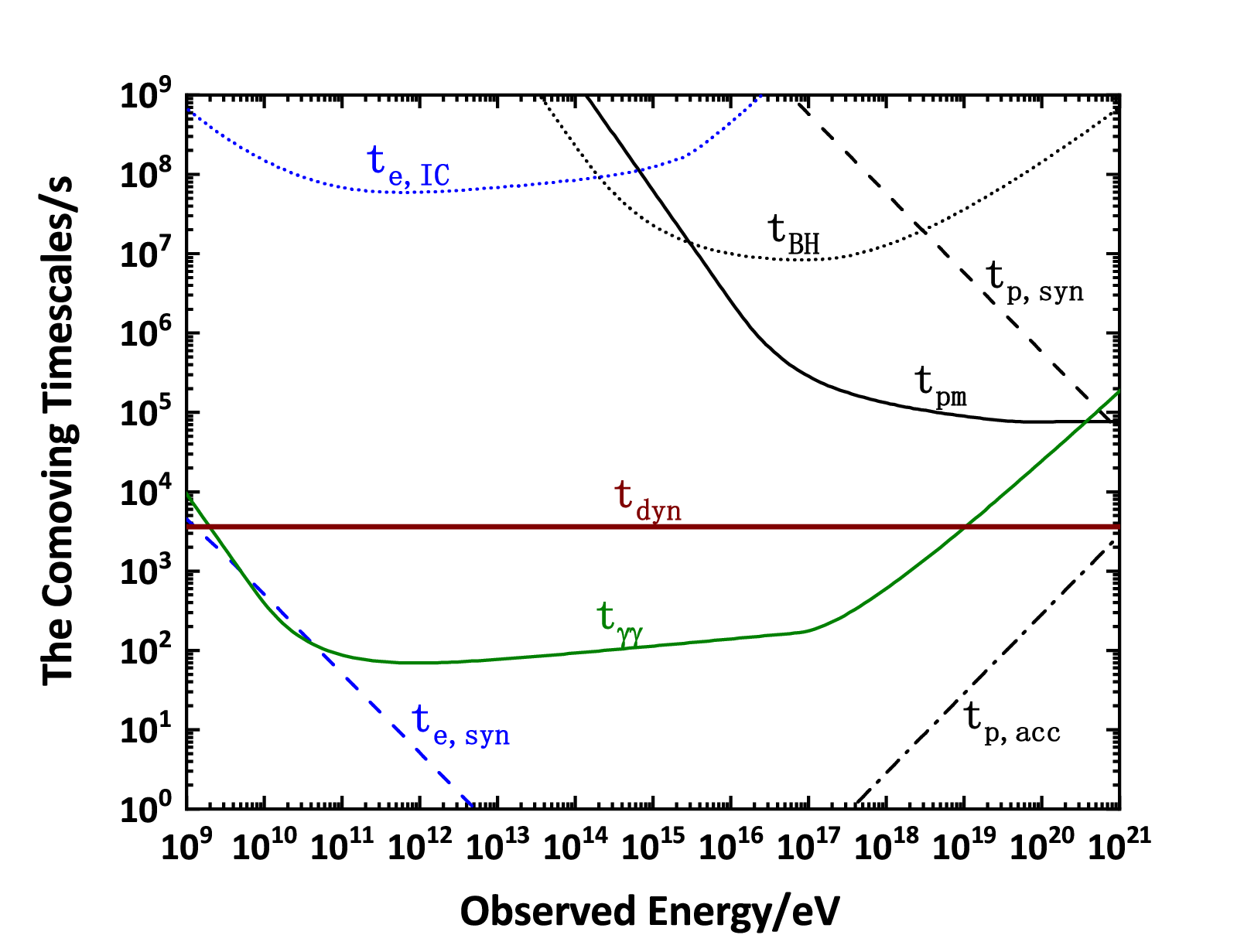} 
	\caption{
		%\textbf{Trajectories of CRs inside a cluster}
		\textbf{Timescales for diverse processes} in the external shocked region at $T=235\,\rm s$ after Fermi-GBM trigger. The magnetic energy factor in the external shocked region is $f_{\rm B,ex}=0.1$, and the averaged keV/MeV spectrum (including an SSA frequency in the low-energy part) with GECAM observations is adopted as the same as in the time interval $231-240\,\rm s$.}
	\label{fig:timescales}
\end{figure}

In order to study the very early afterglow phase, we split the keV/MeV and GeV/TeV time-averaged spectra obtained from the GECAM, Fermi-LAT, and LHAASO into four time intervals, i.e., $231-240\,\rm s$, $240-248\,\rm s$, $248-326\,\rm s$, and $326-900\,\rm s$ after the Fermi-GBM trigger, following the same time interval as in Ref.~\cite{doi:10.1126/science.adg9328}. The synchrotron radiations of non-thermal electrons are invoked to be responsible for the keV/MeV observations in the internal shock region with a large magnetic energy factor $f_{\rm B,in}=30$. Then these keV/MeV photons catch up with the external shocked region with a typical magnetic energy factor $f_{\rm B,ex}=0.1$ and become the target photon field of hadronic processes and EM cascades of secondary high-energy gamma-ray photons and electrons produced by hadronic processes. The timescales for diverse processes in the external shocked region are shown in Fig.~\ref{fig:timescales}. Basically, protons can be accelerated to be UHECRs. In addition, the EM cascade can be fully developed due to the short timescales of electron synchrotron radiations and the $\gamma \gamma $ annihilation. We solve the distribution of cascaded electrons in the quasi-steady state following the treatment in Ref.~\cite{2018ApJ...857...24W}, i.e.,
\begin{equation}
{{n}_e}({{\gamma}_e}) = -\frac{1}{{{\dot{\gamma}}_e}}\int_{{{\gamma }_e}}^\infty  d {{\tilde \gamma }_e}[{{Q_e}({{\tilde \gamma}_e}) + {\dot n}_{e,\gamma \gamma } ({{\tilde \gamma }_e})}],
\label{equne}
\end{equation}
with the electron cooling rate via synchrotron radiation and IC scattering 
\begin{equation}
{{\dot \gamma }_e} =  - \frac{{c{\sigma _T}{{B_{\rm ex}}^2}}}{{6\pi {m_e}{c^2}}}{{\gamma }_e}^2 + {{\dot \gamma}_{e,\rm IC}},
\end{equation}
where, ${{\dot \gamma }_{e,\rm IC}}$ is the IC cooling rate, which can be calculated following \cite{RevModPhys.42.237}.  ${{Q_e}({{ \gamma }_e})}$ is the injection rate of the first-generation electrons (including both $e^-$ and $e^+$) produced by the photomeson production process and the BH process, which is calculated based on the semi-analytical treatment in Ref.~\cite{2008PhRvD..78c4013K}. ${{{\dot n}_{e,\gamma \gamma }}({{\tilde \gamma }_e})}$ is the pair production rate through the $\gamma\gamma$ annihilation, including the annihilation of the high-energy photon from the neutral pion decay produced in the photomeson production process, and the high-energy photon produced by the synchrotron radiation and the IC scattering of high-energy electrons, i.e.,
\begin{equation}
{{\dot n}_{e,\gamma \gamma }}({{\gamma }_e}) = {f_{\rm abs}}({\varepsilon _1})(\dot n_{{\varepsilon _1}}^0 + \dot n_{{\varepsilon _1}}^{\rm syn} + \dot n_{{\varepsilon _1}}^{\rm IC}) + {f_{\rm abs}}({\varepsilon _2})(\dot n_{{\varepsilon _2}}^0 + \dot n_{{\varepsilon _2}}^{\rm syn} + \dot n_{{\varepsilon _2}}^{\rm IC}),
\label{ggab}
\end{equation}
with
\begin{equation}
{f_{\rm abs}}(\varepsilon ) = 1 - \frac{{1 - {e^{ - {\tau _{\gamma \gamma }}(\varepsilon )}}}}{{^{{\tau _{\gamma \gamma }}(\varepsilon )}}}
\end{equation}
being the absorption factors of photons. The $\gamma\gamma$ annihilation produces one electron and one positron, carrying a fraction $f_\gamma$ and $1-f_\gamma$ of the energy of the initial high energy photon, respectively. Based on Ref.~\cite{Bottcher_2013}, $f_\gamma=0.9$ is a good agreement with the numerical Monte Carlo simulations. Therefore, to produce an electron (or positron) with energy ${{\gamma}_e}$, the photon should have the energy of either ${\varepsilon _1} = {{\gamma}_e}/{f_\gamma}$ or ${\varepsilon _2} = {{\gamma }_e}/(1 - {f_\gamma })$. In Eq.~\ref{ggab}, $\dot n_{{\varepsilon }}^0$ is the first-generation photons produced by the neutral pion decay, which can be found by the analytical treatment in Ref.~\cite{2008PhRvD..78c4013K}. Besides, the high-energy photon production rate by the synchrotron radiation $\dot n_{{\varepsilon}}^{\rm syn}$ can be described as
\begin{equation}
\dot n_\varepsilon ^{\rm syn}{\text{ = }}{A_0}{\varepsilon ^{ - 2/3}}\int_1^\infty  {d{{\gamma}_e}} {{n}_e}({{\gamma }_e}){\gamma }_e^{ - 2/3}{e^{ - \varepsilon /(b {\gamma }_e^2)}},
\end{equation}
with
\begin{equation}
{A_0} = \frac{{c{\sigma _T}{{B_{\rm ex}}^2}}}{{6\pi {m_e}{c^2}}}\frac{1}{{\Gamma (4/3){b^{4/3}}}},
\end{equation}
where $\Gamma (4/3) = {\text{0}}{\text{.89297}}$, $b = B_{\rm ex}/{B_{\rm crit}}$ and ${B_{\rm crit}} = 4.4 \times {10^{13}}\,\rm G$, and the high-energy photon production rate by the IC scattering $\dot n_{{\varepsilon }}^{\rm IC}$ can be given by
\begin{equation}
\dot n_\varepsilon ^{\rm IC} = \int_1^\infty  {d{{\gamma }_e}} {{n}_e}({{\gamma }_e})\frac{1}{{{{\gamma }_e}{m_e}{c^2}}}\frac{{dN}}{{dtd{E_1}}},
\end{equation}
where $\frac{{dN}}{{dtd{E_1}}}$ is given by Equation (2.48) in Ref.~\cite{RevModPhys.42.237}.

Since the electron spectrum ${{n}_e}({{\gamma}_e})$ exists on both sides of Eq.~\ref{equne}, this equation can be evaluated progressively, starting from the highest electron energies and then using the solution of ${{n}_e}({{\gamma}_e})$ for large $\gamma_e$ as one progresses toward the lower values of $\gamma_e$, to obtain the equilibrium pair distribution ${{n}_e}({{\gamma}_e})$, which has an excellent agreement with the results of the Monte-Carlo simulations \cite{Bottcher_2013}. Then, one can obtain the synchrotron and IC spectra from the equilibrium pair distribution after the absorption by the target photon field. Since the cascade emission can contribute as the target photon as well, we execute an iteration procedure until the self-consistent results after photon-photon absorption are reached.

We consider the temporal evolution of bulk Lorentz factor and dissipation radius as Eq.~\ref{Eqgamma} and Eq.~\ref{Eqradius}, respectively. For four time intervals, i.e., $231-240\,\rm s$, $240-248\,\rm s$, $248-326\,\rm s$, and $326-900\,\rm s$, we adopt a time bin $1\,\rm s$ to calculate the bulk Lorentz factor and dissipation radius. In the same time interval, for simplification, the averaged keV/MeV spectra derived by GECAM/Fermi-GBM observations are respectively involved. Finally, the cascade emission can be averagely obtained for each time interval and responsible for the observations of Fermi-LAT and LHAASO.

For the afterglow emission in Region II, we implement a detailed simulation based on the description in the section of \textit{``calculation of afterglow model"}. The isotropic kinetic energy $E=10^{55}\,\rm erg$, initial bulk Lorentz factor $\Gamma_0=260$, and wind parameter $A=10^{35.5}\,\rm cm^{-1}$ are fixed as the same as for Region I, while other parameters are determined by the MCMC simulation of the afterglow model and listed in Table~\ref{table:parameter}. Then, the synchrotron and SSC emissions are integrated for four time intervals.

%For a longer time, the prompt keV/MeV radiations become weaker than the afterglow, and the afterglow emission starts to dominate. For these four time intervals at the relatively early stage of the afterglow phase, we adopt a temporal evolutional bulk Lorentz factor and dissipation radius after the GRB jet deceleration, i.e., $\Gamma  = 500{(t/18\,\rm s)^{ - 1/4}}$ and $R = 3 \times {10^{16}}{(t/18\,\rm s)^{1/2}}{\rm{cm}}$ with the time $t=T-T^*$. Before the GRB jet significantly decelerates, i.e., $t<18\,\rm s$, we simply adopt the same bulk Lorentz factor and dissipation radius, i.e., $\Gamma  = 500$ and $R = 3 \times {10^{16}}{\rm{cm}}$. The timescales for diverse processes are shown in Fig.~\ref{fig:timescales}.

\subsection*{Modeling of TeV light curve}

In order to explore the TeV light curve, we implement the GECAM data analyses for diverse time bins, which are selected as small as 1 second for the fast variability period and larger for the relatively smooth period. For each time bin, the photon indexes, the peak energy, and the energy flux in $1\,\mathrm{keV}-10\,\mathrm{MeV}$ are derived and used in the numerical calculations. The energy flux for each time bin is marked in Fig.~\ref{fig:light curve}. We consider two main dissipation regions, which are respectively dominant by prompt photons (Region I) and afterglow emissions (Region II) at the keV/MeV energy band. For Region I, we calculate the hadron-initiated EM cascades, as well as the EM cascades initiated by the electron external inverse Compton (EIC). Protons and electrons are accelerated by the external forward shock and the prompt emission is treated as the seed photons. The EM cascade calculation is the same as in the above section, while the seed photons are adopted by the new and small time bins derived by the GECAM data analyses.

Different from spectral modelings in which the proton luminosity is treated as a free parameter, here, we employ a temporal evolutional proton luminosity. A temporal broken power-law distribution of proton luminosity, i.e.,
\begin{equation}
{L_p} = \left\{ \begin{array}{ll}
{L_{p,c}}{(t/18\,\mathrm{s})^{{\alpha _1}}} & t < 18{\rm{s}}\\
{L_{p,c}}{(t/18\,\mathrm{s})^{{\alpha _2}}} & t \ge 18{\rm{s}}
\end{array} \right.
\end{equation}
is phenomenologically adopted considering the injection is dominant ($\alpha _1  >  0$) before $t_{\rm peak}$ and the cooling is dominant ($\alpha _2  <  0$) after  $t_{\rm peak}$. For simplification, the same distribution form of electrons is adopted but with a smaller critical electron luminosity, $L_{e,c}=L_{p,c} m_e/m_p $ in Region I.

In Region II, we implement a detailed temporal evolution of TeV emission from the afterglow model based on the descriptions in the next section. The parameters are presented in Table~\ref{table:parameter}. 

\subsection*{Calculations of afterglow model}
Considering only the existence of forward shock, a series of papers give their results. We quote here the dynamics of the forward shock of the jet described as
\cite{Nava_2013_Sironi_mnras_v433.p2107..2121,Zhang_2018pgrb.book.....Z},
\begin{equation}\label{eq:dGamma}
\resizebox{0.45\textwidth}{!}{$
\dfrac{d\Gamma}{dr}=
-\dfrac{\Gamma(\Gamma^2-1)(\hat{\gamma}\Gamma-\hat{\gamma}+1)\frac{dm}{dr}c^2
-(\hat{\gamma}-1)\Gamma(\hat{\gamma}\Gamma^2-\hat{\gamma}+1)(3U/r)}
{\Gamma^2[m_0+m]c^2+(\hat{\gamma}^2\Gamma^2-\hat{\gamma}^2+3\hat{\gamma}-2)U}
$},
\end{equation}
\begin{equation}\label{eq:diff_U}
\frac{dU}{dr}=(1-\epsilon)(\Gamma-1) c^{2}\frac{dm}{dr}
-(\hat{\gamma}-1)\left(\frac{3}{r}-\frac{1}{\Gamma} \frac{d \Gamma}{d r}\right) U,
\end{equation}
where ${dm}/{dr}=4 \pi r^{2} n(r) m_p$
with $n(r)$ being the particle density of the circum-burst medium and $m_p$ being the proton mass, and $\Gamma(r)$, $m(r)$, $U(r)$, and $\epsilon$ are the bulk Lorentz factor, the swept-up mass, the internal energy, and the radiation efficiency of electrons in the external-forward shock, respectively. The radiation efficiency has been affected by radiation processes and will be described below. The adiabatic index $\hat{\gamma}$ has been taken from Ref.~\cite{PeEr_2012__apjl_v752.p8..11}.

We denote the instantaneous electron distribution as ${dN_e}/d\gamma_e^{\prime}$,
of which the evolution can be solved based on the continuity equation of electrons
\cite{Fan_2008_Piran_mnras_v384.p1483..1501},
\begin{equation}
\frac{\partial}{\partial r}\left(\frac{dN_e}{d\gamma_e^{\prime}}\right)
+\frac{\partial}{\partial\gamma_e^{\prime}}
\left[\frac{d\gamma^{\prime}_e}{dr}\left(\frac{dN_e}{d\gamma^{\prime}_e}\right)\right]
= Q\left(\gamma^{\prime}_e,r\right)~,
\end{equation}
where ``$\prime$" marks the co-moving frame of shock, and
\begin{equation}
\frac{d\gamma^{\prime}_e}{dr}=-\frac{\sigma_{\rm T}}{6 \pi m_e c^2}
\frac{B^{\prime 2}}{\beta \Gamma}\left[1+Y\left(\gamma^{\prime}_e\right)\right]
{\gamma^{\prime}_e}^2-\frac{\gamma^{\prime}_e}{r}
\end{equation}
is the cooling term, where $\beta=\sqrt{1-\Gamma^{-2}}$, and $\sigma_{\rm T}$ is the Thomson scattering cross section. 
Then, the magnetic field behind the forward shock
is $B^{\prime}={[32\pi { \epsilon_B}{n(r)}]^{1/2}}\Gamma c$,
the Compton parameter
\begin{equation}\label{eq:Y}
Y(\gamma_e^{\prime})=\frac{-1+\sqrt{1+4 \epsilon_{\rm rad} \eta_{\rm KN} \epsilon_e / \epsilon_B}}{2}
\end{equation}
has been solved based on the work of \cite{Fan_2006_Piran_mnras_v369.p197..206} in their appendix, $\epsilon_e$ and $\epsilon_B$ are the equipartition factors for the energy in electrons and the magnetic field in the shock, respectively. $\eta_{\rm KN}$ is suppression factor due to the Klein-Nishina effect.
Additionally, one can have
$\epsilon=\epsilon_{\rm rad}\epsilon_e$
with $\epsilon_{\rm rad}=\min \{1,(\gamma^{\prime}_m/\gamma^{\prime}_c)^{(p-2)}\}$
\cite{Sari_2001_Esin_apj_v548.p787..799,Fan_2008_Piran_mnras_v384.p1483..1501}, with $\gamma^{\prime}_c=6 \pi m_e c/[\sigma_{\rm T}\Gamma {B'}^2 t'(1+Y)]$ and $\gamma^{\prime}_m=\epsilon_e/\xi_e(p-2)m_p\Gamma/[(p-1)m_e]$
\cite{Sari_1998_Piran_apj_v497.p17..20L},
where we assume that a fraction $\xi_e$ of the electron population is shock-accelerated \cite{2005ApJ...627..861E}. The swept-in electrons by the shock are accelerated to a power-law distribution of Lorentz factor $\gamma^{\prime}_e$,
i.e., $Q\propto {\gamma^{\prime}_e}^{-p}$ for
$\gamma^{\prime}_m
\leqslant \gamma^{\prime}_e \leqslant \gamma^{\prime}_M$,
where $p (>2)$ is the power-law index, and 
$\gamma^{\prime}_M=\sqrt{9m_e^2{c^4}/[8B'{e}^3(1+Y)]}$ with $e$ being the electron charge
\cite{Kumar_2012_Hernandez_mnras_v427.p40..44L}.
Evidently, the overall inflow of nonthermal electrons should be
\begin{equation}\label{eq:Qinj}
\int Q\left(\gamma^{\prime}_e,r\right) d\gamma^{\prime}_e
=4 \pi r^2  n(r) \xi_e dr.
\end{equation}

The emitted spectral power of synchrotron radiation
at a given frequency $\nu'$ of a single electron is
\begin{equation}
P^{\prime}_e(\nu^{\prime}, \gamma^{\prime}_e)
=\frac{\sqrt{3} q_e^3 B^{\prime}}{m_e c^2}
F\left(\frac{\nu'}{\nu'_c}\right),
\end{equation}
where
$F(x)=x\int_{x}^{+\infty}K_{5/3}(k)dk$,
$K_{5/3}(k)$ is the modified Bessel function of $5/3$ order,
and $\nu'_{\rm c}=3q_e B'{\gamma'_e}^2/(4\pi m_e c)$, respectively.
Thus, the spectral power of synchrotron radiation of electrons
${dN_e}/d\gamma_e^{\prime}$ at a given frequency $\nu'$ is
\begin{equation}
P'_{\rm syn}(\nu')=
\int_{\gamma^{\prime}_m}^{\gamma^{\prime}_M}
P'_e(\nu')\frac{dN_e}{d\gamma_e^{\prime}}d\gamma^{\prime}_e,
\end{equation}
thus the synchrotron photon spectra are
$n'_{\gamma,\rm syn}\left(\nu^{\prime}\right) \simeq
\frac{P'_{\rm syn}(\nu')}{4\pi r^2 c h\nu'}$.

The emission of the SSC process is calculated based on the electron spectrum
${dN_e}/d\gamma_e^{\prime}$ and target seed photons of synchrotron radiation
$n_{\gamma,\rm syn}^{\prime}\left(\nu_t^{\prime}\right)$
from the synchrotron radiation
\cite{Fan_2008_Piran_mnras_v384.p1483..1501}
\begin{equation}
P_{\rm SSC}^{\prime}\left(\nu^{\prime}\right)
=\frac{3 \sigma_{\rm T}ch\nu^{\prime}}{4}
\int_{\nu_{\min}^{\prime}}^{\nu_{\max}^{\prime}}
\frac{n_{\gamma, \rm syn}^{\prime}\left(\nu_t^{\prime}\right) d\nu_t^{\prime}}{\nu_t^{\prime}}
\int_{\gamma_m^{\prime}}^{\gamma_M^{\prime}} \frac{F(q, g)}{\gamma_e^{\prime 2}}
\frac{dN_e}{d\gamma_e^{\prime}} d\gamma_e^{\prime},
\end{equation}
where
$F(q, g)=2q\ln q+(1+2q)(1-q)+8q^2 g^2(1-q)(1+4qg)$,
$q=w/4g(1-w)$,
$g=\gamma_e^{\prime}h\nu_t^{\prime}/m_e c^2$,
and $w=h\nu^{\prime}/\gamma_e^{\prime} m_e c^2$, respectively.
One can also derive the SSC photon spectra $n'_{\gamma,\rm SSC}$.
The total spectral power and photon spectra then be
$P'_{\rm tot}(\nu')=P'_{\rm syn}(\nu')+P'_{\rm SSC}(\nu')$
and
$n'_{\gamma,\rm tot}=n'_{\gamma,\rm syn}+n'_{\gamma,\rm SSC}$,
respectively.

The cross-section of $\gamma\gamma$ annihilation reads\cite{Gould_1967_Schreder_PhRv..155.1404}
\begin{equation}
\sigma_{\gamma\gamma}=\frac{3\sigma_{T}}{16}\left(1-\beta_{\rm cm}^2\right)
\left[2\beta_{\rm cm}^3-4\beta_{\rm cm}+\left(3-\beta_{\rm cm}^4\right)
\ln \frac{1+\beta_{\rm cm}}{1-\beta_{\rm cm}}\right],
\end{equation}
where
$\beta_{\rm cm}=\sqrt{1-2\left(m_e c^2\right)^2/\left[h\nu^{\prime} h\nu^{\prime}_{t}(1-\cos{\theta})\right]}$, $\theta$ is the angle of collision photons, and $\nu^{\prime}_{t}$ is the frequency of target photons, respectively.
It is obvious that there is a physical value only when
$h\nu^{\prime} h\nu^{\prime}_t(1-\cos{\theta}) \geqslant 2\left(m_e c^2\right)^2$.
Thus, the optical depth for a gamma-ray photon with frequency $\nu^{\prime}$
is expressed as \cite{Murase_2011_Toma_apj_v732.p77..77}
\begin{equation}
\tau^{\gamma\gamma}(\nu^{\prime})=\frac{\Delta}{2} \int_{-1}^{1} (1-\mu)d\mu
\int\sigma_{\gamma\gamma} n'_{\gamma,\rm tot}(\nu^{\prime}_t) d\nu^{\prime}_t,
\end{equation}
where $\Delta=r/12\Gamma$. 

The synchrotron self-absorption (SSA) effect has also been considered
in our numerical calculations. The optical depth is the function of both the electron distribution and synchrotron radiation power of a single electron,
\begin{equation}
\tau^{\rm SSA}(\nu^{\prime})=-\frac{\Delta}{8 \pi m_e \nu^{\prime 2}}
\int \gamma^{\prime 2}_e P^{\prime}_e(\nu^{\prime}, \gamma^{\prime}_e)
\frac{\partial}{\partial \gamma^{\prime}_e}
\left[\frac{dN_e/d\gamma_e^{\prime}}{\gamma_e^{\prime 2}}
\right] d\gamma_e^{\prime}.
\end{equation}

Finally, the intrinsic spectral power of afterglow is
\begin{equation}
P^{\prime \rm in}_{\rm tot}(\nu^{\prime})=
P^{\prime}_{\rm tot}(\nu^{\prime})
\frac{1-e^{-\left[\tau^{\gamma\gamma}(\nu^{\prime})
+\tau^{\rm SSA}(\nu^{\prime})\right]}}
{\tau^{\gamma\gamma}(\nu^{\prime})
+\tau^{\rm SSA}(\nu^{\prime})}.
\end{equation}

Besides, GRBs are believed to be generated by
the motion of ultra-relativistic jets within a half-opening angle $\theta_j$.
In this work, we set the GRB jet as an on-axis-observed jet, i.e., $\theta_v=0$.
Assuming the observed frequency is $\nu_{\rm obs}$, the frequency transformed to the co-moving frame is denoted as
$\nu^{\prime}(\nu_{\rm obs})=(1+z)\nu_{\rm obs}/\mathcal{D}$, where $z$ is the redshift,
$\mathcal{D}=1/\left[\Gamma(1-\beta\cos\Theta)\right]$
is the Doppler factor,
and $\Theta$ represents the angle between the direction of motion
of a fluid element in the jet and the line of sight.
When taking the equal-arrival-time surface (EATS)
effect into account \cite{Waxman_1997_ApJ...485L...5W}, the observed intrinsic spectral flux should be
\begin{equation}
F^{\rm in}_{\rm tot}\left(\nu_{\rm obs}\right)=
\frac{1+z}{4 \pi D_{L}^2} \int_{0}^{\theta_j}
P^{\prime \rm in}_{\rm tot}\left[\nu^{\prime}\left(\nu_{\rm obs}\right)\right]
\mathcal{D}^3 \frac{\sin \theta}{2} d\theta,
\end{equation}
where $D_L$ is the luminosity distance from the burst to the Earth.
The integration of $\theta$ is performed over an elliptical surface
that photons emitted from the surface have the same arrival time for an observer
\cite{Geng_2018_Huang_apjs_v234.p3..3},
\begin{equation}
t_{\mathrm{obs}}=
(1+z) \int \frac{1-\beta\cos\Theta}{\beta c} dr \equiv \rm{const}.
\end{equation}

\section*{Data availability}
The GECAM data that support the plots within this paper and other findings of this study will be available upon request and will be released on the GECAM website (http://gecam.ihep.ac.cn/). The Fermi-GBM and Fermi-LAT data are publicly available at the Fermi Science Support Center (http://fermi.gsfc.nasa.gov/ssc/data/access/). The LHAASO data can be read from their paper \cite{doi:10.1126/science.adg9328}.

\section*{Author contributions}
K.W. initiated and led the project, performed the theoretical spectral and light curve modelings, and writing of the original manuscript; Q.W.T. performed the Fermi-GBM and Fermi-LAT data analysis, the correlation analysis between the Fermi-LAT and LHAASO data, and writing; Y.Q.Z., C. Z., and S.L.X. implemented the GECAM data analysis and writing; J.R. calculated the radiations of afterglow model, and writing; B.Z. provided advice and expertise while interpreting observational data, and improved the writing of the manuscript. 

\section*{Competing interests}
The authors declare no competing interests.

\section*{Acknowledgements}
K.W. acknowledges support from the National Natural Science Foundation of China under grant No.12003007 and the Fundamental Research Funds for the Central Universities (No. 2020kfyXJJS039). Q.W.T. acknowledges support from the National Natural Science Foundation of China under grant No.12065017 and the Natural Science Foundation of Jiangxi Province of China grant No. 20224ACB211001. S.L.X. acknowledges support from the National Natural Science Foundation of China under grant No.12273042. The GECAM (Huairou-1) mission is supported by the Strategic Priority Research Program on Space Science (Grant No. XDA15360000, XDA15360102, XDA15360300) of the Chinese Academy of Sciences.

\bibliography{reference}

%\bibliography{sample631}{}
%\bibliographystyle{aasjournal}

% \appendix
% % \section{Mean Free Paths }\label{Appsec:MFP}

\end{document}